\documentclass[conference,letterpaper]{IEEEtran}

\usepackage[utf8]{inputenc}
\usepackage[T1]{fontenc}
\usepackage{url}
\usepackage{ifthen}
\usepackage{cite}
\usepackage[cmex10]{amsmath} % Use the [cmex10] option to ensure complicance
\usepackage{graphicx}
\usepackage{latexsym}
\usepackage{amsthm}
\usepackage{ascmac}
\usepackage{xcolor}
\usepackage{enumitem}
\newtheorem{theorem}{Theorem}
\newtheorem{definition}{Definition}
\newtheorem{remark}{Remark}
\newtheorem{lemma}{Lemma}

\renewcommand{\figurename}{Fig.}
\renewcommand{\refname}{References}

\newcommand{\defeq}{\mathrel{\mathop:}=}

\begin{document}
\title{Fundamental Limits of Identification System
With Secret Binding Under Noisy Enrollment}

% %%% Single author, or several authors with same affiliation:
% \author{%
%   \IEEEauthorblockN{Stefan M.~Moser}
%   \IEEEauthorblockA{ETH Zürich\\
%                     ISI (D-ITET)\\
%                     CH-8092 Zürich, Switzerland\\
%                     Email: moser@isi.ee.ethz.ch}
% }

%%% Several authors with up to three affiliations:
\author{%
  \IEEEauthorblockN{Vamoua Yachongka}
  \IEEEauthorblockA{Dept. of Computer and Network Engineering\\
                    The University of Electro-Communications\\
                    Tokyo, Japan\\
                    Email: va.yachongka@uec.ac.jp}
  \and
  \IEEEauthorblockN{Hideki Yagi}
  \IEEEauthorblockA{Dept. of Computer and Network Engineering\\
                    The University of Electro-Communications\\ 
                    Tokyo, Japan\\
                    Email: h.yagi@uec.ac.jp}
}

\maketitle

%%%%%%
%% Abstract: 
%% If your paper is eligible for the student paper award, please add
%% the comment "THIS PAPER IS ELIGIBLE FOR THE STUDENT PAPER
%% AWARD." as a first line in the abstract. 
%% For the final version of the accepted paper, please do not forget
%% to remove this comment!
%%
\begin{abstract}
We study fundamental limits of biometric identification systems with chosen secret
from an information theoretic perspective.
Ignatenko and Willems (2015) characterized the capacity region of identification, secrecy, and privacy-leakage rates
of the system provided that {the enrollment channel} is noiseless.
In the enrollment process, however,
it is highly considered that noise occurs when bio-data is scanned.
Recently, Yachongka and Yagi (2019) characterized the capacity region of
a different system (generated secret system) considering noisy enrollment and template constraint.
In this paper, we are interested in characterizing
the capacity region of identification, secrecy, template, and privacy-leakage rates of
the system with chosen secret under the same settings as Yachongka and Yagi (2019).
As special cases, the obtained result shows that
the characterization reduces to the one given by Ignatenko and Willems (2015)
where the enrollment channel is noiseless and there is no constraint on the template rate, and
it also coincides with the result derived by G\"unl\"u and Kramer (2018) where
there is only one individual.
\end{abstract}

\section{Introduction}

Biometric {identification} is a process of comparing biological characteristics or data
(bio-data) of an {individual} to that individual's template
{already stored in the system database.
%The aim is to recognize the individual by matching the unique features with the
%templates enrolled in the system database
%where several other individuals' information {kept as} well.
Some well-known applications are fingerprint-based identification, iris-based identification,
voice recognition, etc.
%Nowadays many applications make use of
%this technology like homeland checking at land port, mobile payment with smart phone, etc.

O'Sullivan and Schmid \cite{OS} and Willems et al.\ \cite{willems} separately introduced the
discrete memoryless {\em biometric identification system} (BIS).
Willems et al.\ \cite{willems} have clarified
the identification capacity of the BIS, which is the maximum achievable rate of the number of
individuals when the error probability converges to zero as the length of biometric
data sequences goes to infinity.
However, the implementation in \cite{willems} stores
bio-data {sequences in} the system database in a plain form, leading
to a critical privacy-leakage threat. Later,
Tuncel \cite{tuncel} has developed their model by
incorporating compression of bio-data {sequences and}
clarified the capacity region of identification and coding rates
(In this study, a codeword or helper data is called a template, and this coding rate is called the template rate).

Besides, there are several studies dealing with {the} secrecy rate of the BIS. An example of them
is {the} BIS model with chosen secret. In this model, secret key or key of individual is
chosen independently of bio-data and we name {the} BIS model with chosen secret as
the CS-BIS model.
%In {the} CS-BIS model, the given secret key
%is chosen independently of {bio-data}.
Ignatenko and Willems \cite{itw1} and
Lai et al.\ \cite{lhp} investigated the fundamental trade-off between
secret key and privacy-leakage rates {in the CS-BIS model}.
Ignatenko and Willems \cite{itw3} extended the model studied in \cite{itw1} to consider individual's
estimation and characterized
the capacity region of identification, secrecy, and privacy-leakage rates in the
CS-BIS model. One thing to be noted is that {all three studies} mentioned above
{assumed} that {the enrollment channel is noiseless}.
However, when bio-data is scanned, it is highly considerable that bio-data sequences are
subject to noise so it is important to consider {the enrollment channel is noisy} like
{in \cite{willems}, \cite{tuncel}, \cite{onur}, and \cite{vy3}}.
Related to the studies on {the} CS-BIS model, a generated secret BIS (GS-BIS) model,
{where}
secret key is extracted from bio-data sequence{,} is studied in \cite{itw3}--\hspace{-0.1mm}\cite{vy1}}.

Studies on the CS-BIS model without estimating individual are extensively discussed in, e.g.
\cite{itw1}, \cite{lhp}, \cite{KY}, and \cite{onur}.
%and {incorporating}
%the analysis of {the} template rate {as well as the}
%secrecy and privacy-leakage rates can be found in \cite{KY} and \cite{onur}.
More precisely,
Koide and Yamamoto {\cite{KY}} analyzed the model for non-negligible secrecy-leakage.
G\"unl\"u and Kramer \cite{onur} evaluated the model by treating the enrollment channel
is noisy ({\em hidden} source model). The benefit of having single
{individual} made a successful breakthrough for them to prove the capacity region
by {\em one} auxiliary random variable (RV) in an elegant way. When the model is extended to the one
with considering {individual}'s estimation,
it seems difficult to use the same analyzing techniques, especially the evaluating of privacy-leakage rate,
to characterize the capacity region. {In the GS-BIS scenario, Yachongka and Yagi \cite{vy3}
have characterized the capacity region of the model with {{\em two}} auxiliary RVs recently.
Then, an interesting question is if the same arguments also work for {the CS}-BIS scenario.}

In this paper, we aim to characterize the capacity region of identification, secrecy, template,
and privacy-leakage rates for the CS-BIS model. {Compared to the model proposed in \cite{itw3}}, we analyze the region under conditions that
\begin{enumerate}[label=\arabic*)]
	\item adding {\em noisy} enrollment channel,
	\item constraining template rate,
	\item assuming that the prior distribution of the identified individual is unknown.
\end{enumerate}
We show that it is possible to characterize the capacity region of the CS-BIS model in two different ways.
A characterization uses a single auxiliary RV and another requires
two auxiliary RVs. In this scenario, we will prove the capacity region based on the latter by applying
the technique developed in \cite{vy3}.
As special cases, it can be checked that our characterization reduces to the one given by
Ignatenko and Willems \cite[Theorem 2]{itw3}
where the enrollment channel is noiseless and there is no constraint on the template rate, and
it also corresponds to the result derived by G\"unl\"u and Kramer \cite[Theorem 2]{onur} where there is
no consideration of individual's estimation.

The rest of this paper is organized as {follows}. In Sect.\ \ref{sec2},
we define notation used in this paper and describe the details of the system model.
In Sect.\ \ref{sec3}, {we present} the problem formulation and main result. Next,
we highlight the proof of the main result in Sect.\ \ref{sec4}. Finally,
in Sect.\ \ref{sec5}, we give some concluding remarks and future works.

\section{Notation and Model Descriptions} \label{sec2}
%\vspace{-2mm}
%In this section, we define notation used in this paper and describe the details
%of the system model.
% model from information theoretic framework.
%\vspace{-2mm}
\subsection{Notation}
%\vspace{-2mm}
Calligraphic $\mathcal{A}$ stands for a finite alphabet. Upper-case $A$ denotes a {RV}
taking values in
$\mathcal{A}$ and lower-case $a \in \mathcal{A}$ denotes its realization.
$P_A(a)~\defeq~\Pr[A = a]$, $a \in \mathcal{A}$, represents 
{the} probability distribution on $\mathcal{A}$, and $P_{A^n}$ represents {the} probability distribution 
of RV $A^n = (A_1,\cdots,A_n)$ in $\mathcal{A}^n$, the $n$-th Cartesian product of 
$\mathcal{A}$. $P_{A^nB^n}$ represents the joint probability distribution of a pair of
RVs $(A^n,B^n)$ and its conditional probability distribution
$P_{A^n|B^n}$ is defined as
\begin{align}
&P_{A^n|B^n}(a^n|b^n) = \frac{P_{A^nB^n}(a^n,b^n)}{P_{B^n}(b^n)}\nonumber \\
&~~~~(\forall a^n \in \mathcal{A}^n, \forall b^n \in \mathcal{B}^n ~\mathrm{such}~\mathrm{that}~P_{B^n}(b^n)~>~0).
\end{align}

The entropy of RV $A$ is denoted by $H(A)$, the joint entropy of RVs $A$ and $B$ 
%which take values in $\mathcal{A}\times \mathcal{B}$
is denoted by
$H(A,B)$, and the mutual information between $A$ and $B$ is denoted by $I(A;B)$ \cite{cover}.
Throughout this paper, logarithms are of base two.
For integers $a$ and $b$ such that
$a < b$, $[a,b]$ denotes the set 
$\{a,a+1,\cdots,b\}$. A partial sequence of a sequence $c^n$ from the first symbol
to the $t$th symbol $(c_1,\cdots,c_{t})$ is represented by $c^{t}$.

A sequence $x^n \in \mathcal{X}^n$ is said to be $\delta$-$strongly~typical$ with
respect to a distribution {$P_{X}$} on $\mathcal{X}$ if $|\frac{1}{n}N(a|x^n) - P_{X}(a)| \leq \delta$ and
$P_{X}(a) = 0$ implies $\frac{1}{n}N(a|x^n) = 0$ for all $a \in \mathcal{X}$, where $N(a|x^n)$
is the number of occurrences of $a$ in the sequence $x^n$, and $\delta$ is
an arbitrary positive number.
The set of sequences $x^n \in \mathcal{X}^n$ such that $x^n$ is $\epsilon$-strongly typical is called
the strongly typical set and is denoted by $A^{(n)}_{\epsilon}(X)$ (cf. \cite{cover}, \cite{GK}). This concept is easily extended to
joint distributions.

\subsection{Model Descriptions}
\begin{figure}[!h]
 \begin{center}
  \includegraphics[width = 85mm]{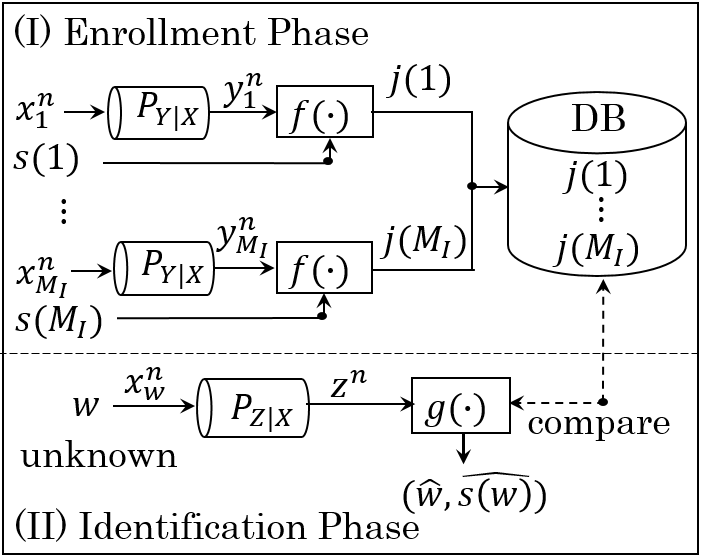}
 \end{center}
 %\vspace{-3mm}
 \caption{CS-BIS model}
 %\vspace{-3mm}
 \label{fig:gdmc}
\end{figure}

The CS-BIS model considered in this paper is shown in Fig.\ \ref{fig:gdmc}.
Basically, it consists of two phases: (I) {\em Enrollment Phase} and
(I\hspace{-.1em}I) {\em Identification Phase}. Next we explain the details of each phase. 
%\vspace{-3mm}
%\vspace{-2mm}

\medskip
\noindent{(}I) {\em Enrollment Phase}:

Let $\mathcal{I} = [1,M_I]$ and $\mathcal{X}$ be the set of indexes of individuals and
a finite source alphabet, respectively.
%We assume that there are $M_I$ individuals whose index takes a value
%in  $\mathcal{I}$.
For any $i \in \mathcal{I}$, we assume that
$x^n_i = (x_{i1},\cdots ,x_{in}) \in \mathcal{X}^n$, an $n$-length
bio-data sequence of individual $i$, is generated %independently and identically distributed
i.i.d. from a stationary memoryless source $P_X$. The generating probability for each 
sequence $x^n_i \in \mathcal{X}^n$ is
%\vspace{-4mm}
\begin{equation}
P_{{X^n_i}}(x^n_i)~\defeq~\Pr[{X^n_i} = x^n_i]~=~\textstyle \prod_{\substack{k = 1}}^nP_{X}(x_{ik}).
\label{dataseisei}
\end{equation}
%\vspace{-5mm}
%\noindent{f}or all $i \in \mathcal{I}$. In this paper, $x^n_i$ is called a bio-data sequence of individual $i$.
%{\em Enrollment Phase}:

Now let $\mathcal{J} = [1, M_J]$ and $\mathcal{S} = [1, M_S]$
be the set of indexes of templates stored in the database and individuals' secret key, respectively.
%In this phase, 
All bio-data sequences are observed via a {\em discrete memoryless channel} (DMC)
$\{\mathcal{Y},P_{Y|X},\mathcal{X}\}$, where
$\mathcal{Y}$ is a finite output-alphabet of $P_{Y|X}$.
The corresponding probability that $x^n_i \in \mathcal{X}^n$
is observed as $y^n_i = (y_{i1},y_{i2},\cdots ,y_{in}) \in \mathcal{Y}^n$ via the DMC $P_{Y|X}$ is
%\vspace{-5mm}
\begin{eqnarray}
P_{Y^n_i|X^n_i}(y^n_i|x^n_i)~=~\textstyle \prod_{\substack{k = 1}}^nP_{Y|X}(y_{ik}|x_{ik}) 
\label{eq2}
\end{eqnarray}
%\vspace{-5mm}

\noindent{f}or all $i \in \mathcal{I}$.

A secret key $s(i) \in \cal{S}$ is chosen uniformly at random and independent of all other RVs.
Encoder mapping $f$ encodes $y^n_i$ and $s(i)$ into template $j(i) \in \cal{J}$ as
%\vspace{-3mm}
$
j(i) = f(y^n_i,s(i))
$. The  template $j(i)$ is stored at position $i$ in the database, which can be accessed
by the decoder.
%the secret key {$s(i)$} is returned to individual $i$ and kept confidential.

\medskip
\noindent{(}I\hspace{-.1em}I) {\em Identification Phase}:

Bio-data sequence $x_w^n\ (w \in\mathcal{I})$ of an unknown $w$
({index of individual has already enrolled} in the database)
is observed via a DMC $\{\mathcal{Z},P_{Z|X},\mathcal{X}\}$, where $\mathcal{Z}$ is a finite output-alphabet of 
$P_{Z|X}$. The probability that $x^n_w \in \mathcal{X}^n$ is output as $z^n = (z_{1},z_{2},\cdots ,z_{n}) \in \mathcal{Z}^n$
via $P_{Z|X}$ is given by
%\vspace{-3mm}
\begin{eqnarray}
P_{Z^n|X^n_w}(z^n|x^n_w)~=~\textstyle \prod_{\substack{k = 1}}^nP_{Z|X}(z_k|x_{wk}). \label{eq3}
\end{eqnarray}
%\vspace{-5mm}

\noindent{T}he sequence $z^n$ is passed to the decoder
$
g: \mathcal{Z}^n \times \mathcal{J}^{M_I} \longrightarrow \mathcal{I} \times \mathcal{S},
$
comparing $z^n$ with templates in the database and 
outputs the pair of estimated value %of an unknown individual's index
$(\widehat{w},\widehat{s(w)})$.
%In other words, an estimated index $\widehat{i}$ is calculated as
%$(\widehat{w},\widehat{s(w)}) = g\left(z^n,j(1),\cdots,j(M_I)\right)$.
%\vspace{-2mm}
\begin{remark}
Note that the distribution of $P_{X}$, $P_{Y|X}$, and {$P_{Z|X}$} are assumed to be known or fixed
and RV $W$ is independent of $(X^n_i,Y^n_i,J(i),S(i),Z^n)$ for all $i \in \mathcal{I}$ like
previous studies. But, in this paper we assume neither that the identified individual
index $W$ are uniformly distributed over $\mathcal{I}$ nor that there is a prior distribution of
{$W$}.
\end{remark}

The motivation {to analyze} performance of the BIS provided that the distribution of
$I$ is unknown is that the identified frequencies of each individual
are likely different. For example, it is hard to think that the frequencies of coming
to use a bank teller of each {individual} are identical.
This assumption is important to take care of from real application perspective.

\section{Definitions and Main Results} \label{sec3}
The formal definition of the addressed problem and the main theorem of this study are given
{below}.
\begin{definition} \label{def1}
A tuple of an identification, secrecy,
template, and privacy-leakage rates $(R_I,R_S,R_J,R_L)$ is said to be achievable if for any $\delta > 0$
and large enough $n$ there exist pairs of encoders and decoders that satisfy for all
${i} \in \mathcal{I}$
\begin{align}
  \max_{\substack{i \in \mathcal{I}}}\Pr\{(\widehat{W},\widehat{S(W)}) &\neq (W,S(W))|W = i\} \leq  \delta, \label{a} \\
  \frac{1}{n}\log{M_I} &\geq R_I - \delta{,} \label{b} \\
  \frac{1}{n}\log{M_J} &\leq R_J + \delta, \label{d} \\
  {\frac{1}{n}\log{M_S}} &\geq R_S - \delta, \label{c} \\
  \max_{\substack{i \in \mathcal{I}}}\frac{1}{n}I(X^n_i;J(i)) &\leq R_L + \delta, \label{dd} \\
  \max_{\substack{i \in \mathcal{I}}}\frac{1}{n}I(S(i);J(i)) &\leq \delta. \label{shiki4}
\end{align}
Moreover, the capacity region $\mathcal{R}$ is defined as the closure of the set of all achievable
rate tuples.
%,called the capacity region.
\end{definition}
In Definition \ref{def1}, (\ref{a}) is the condition of error probability
of individual $i$ which should be arbitrarily small.
Equations (\ref{b}), (\ref{d}) and (\ref{c}) are
the constraints related to identification, template and secrecy rates, respectively.
In terms of the privacy protection perspective, we measure the information leakage of individual $i$ by
(\ref{dd}) and (\ref{shiki4}).
Condition (\ref{dd}) measures the amount of privacy-leakage of original bio-data $X^n_i$
from template {$J(i)$}
in the database and it must be smaller than or equal to {$R_L + \delta$}.
%Later, we will see that the evaluation for $R_L$ is the most {\em intricate task}.
Condition (\ref{shiki4}) measures the secrecy-leakage
between the template and the secret key of individual $i$ and
it requires that {the leaked amount} is arbitrarily small.

\begin{theorem}\label{th1}
The capacity region for the CS-BIS model is given by
\begin{align}
\mathcal{R} = \mathcal{A}_1,
\end{align}
where $\mathcal{A}_1$ is defined as
\begin{align}
\mathcal{A}_1 = \{(R_I,&R_S,R_J,R_L): \nonumber \\
&R_I + R_S \leq I(Z;U),\nonumber \\
&R_J \geq I(Y;U),\nonumber \\
&R_L \geq I(X;U) - I(Z;U) + R_I, \nonumber \\
&R_I \geq 0, R_S \geq 0 \nonumber \\
&\mathrm{for~some}~U~\mathrm{s.t.}~Z-X-Y-U\}, \label{theorem1}
\end{align}
where auxiliary RV $U$ takes values in a
finite alphabet $\mathcal{U}$ with $|\mathcal{U}| \leq |\mathcal{Y}| + 2$.
\qed
\end{theorem}

\begin{remark} \label{remark2222}
We define a region $\mathcal{A}_2$ as
\begin{align}
\hspace{-0mm} \mathcal{A}_2 = \{(R_I,&R_S,R_J,R_L):\nonumber \\
0 \leq~& R_I \leq I(Z;V),\nonumber \\
0 \leq~& R_S \leq I(Z;U)-I(Z;V),\nonumber \\
&R_J \geq I(Y;U),\nonumber \\
&R_L \geq I(X;U) - I(Z;U) + I(Z;V), \nonumber \\
&{\mathrm{for~some}~U~\mathrm{and}~V~\mathrm{s.t.}~ Z-X-Y-U-V}
\}, \label{theorem2}
\end{align}
\noindent{w}here auxiliary RVs $U$ and $V$ take values in some
finite alphabets $\mathcal{U}$ and $\mathcal{V}$
with $|\mathcal{U}| \leq (|\mathcal{Y}| + 2)(|\mathcal{Y}| + 3)$ and $|\mathcal{V}| \leq |\mathcal{Y}|+ 3$.
Then, it can be verified that
\begin{align}
\mathcal{A}_1 = \mathcal{A}_2. \label{region333}
\end{align}
The proof can be done by similar arguments shown in \cite[Appendix A]{vy3} therefore omitted. In this {paper}, we will
{prove} Theorem \ref{th1} based on
the {rate constraints} of the region $\mathcal{A}_2$ instead of $\mathcal{A}_1$.
\end{remark}

\begin{figure*}[!t]
 \begin{center}
  \includegraphics[width = 170mm]{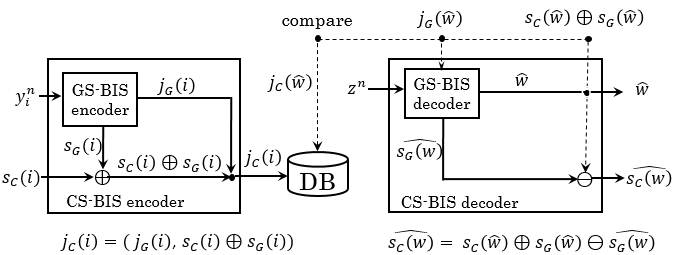}
 \end{center}
 %\vspace{-4mm}
 \caption{Encoder and decoder of CS-BIS model}
 %\vspace{-3mm}
 \label{fig4}
\end{figure*}

\begin{remark}
Likewise the observation in \cite{onur}, the capacity region of GS-BIS model (cf. \cite[Theorem 1]{vy3}) is clearly 
wider than $\mathcal{R}$,
{which is due to the bound} {on} $R_J$.
A remark given in \cite{vy1} indicated that in case {where} the enrollment channel is noiseless $(X=Y)$,
the fundamental limit of $R_L$ and $R_J$ is identical for the GS-BIS model. However,
this claim does not apply to the CS-BIS model.
\end{remark}

One can check that the characterization of Theorem 1 coincides with
the region characterized by Ignatenko and Willems \cite[Theorem 2]{itw3} {in two steps:
first replace $Y$ by $X$ and then remove the constraint $R_J$ from (\ref{theorem1}).
The obtained region {is identical} to
the result in \cite[Theorem 2]{itw3} where the enrollment channel is
noiseless ($X=Y$) and the template rate can be arbitrarily large}.
%The claim can be proved by resembling arguments seen in the proof
%of Remark \ref{remark2222}.
Also, this characterization corresponds to the region given
by G\"unl\"u and Kramer \cite[Theorem 2]{GK} {with only one individual}.
It is easy to check this claim by just
setting $R_I = 0$.

\section{Proof of Theorem \ref{th1}} \label{sec4}
In this section, we only give a guideline of how to prove Theorem \ref{th1}. For detailed proofs,
we recommend the readers refer to check \cite[Proof of Theorem 1]{vy3}.

\subsection{Achievability (Direct) Part}
In order to avoid the confusion in the following arguments, we introduce some new notations which are used only
in this part. The pairs $(J_C(i),S_C(i))$ and $(J_G(i),S_G(i))$ denote the template and the
secret key of individual $i$ for CS-BIS and GS-BIS encoders, respectively.
Moreover, $M_{J_C}$ and $M_{J_G}$ denote the number of templates of the CS-BIS and GS-BIS
models\footnote{Normally,
$J_C(i)$, $S_C(i)$, and $M_{J_C}$ are denoted by $J(i)$, $S(i)$, and $M_J$
in other sections of this paper.}.

\medskip
\noindent{\em Overviews}:

The proof idea of this part is based on the achievability proof of the GS-BIS model provided in \cite{vy3}.
The difference
is that the encoder and decoder of the GS-BIS model are used as components inside the encoder
and decoder of the CS-BIS model as shown in Fig.\ \ref{fig4}.
For encoding in the CS-BIS model,
a so-called masking layer (one-time pad operation) is used
to mask $s_C(w) \in \cal{S}$ for secure transmission by using $s_G(w) \in \cal{S}$
as $s_C(w)\oplus s_G(w)$.
%, where $S_G(i)$ and $J_G(i)$ denote the secret key and
%the template of individual $i$ generated from {the} encoder in the GS-BIS.
The template $j_C(w)$ is the combined information of $j_G(i)$ and the masked data
$s_C(w)\oplus s_G(w)$, i.e.,
\begin{align}
j_C(w) = (j_G(w),s_C(w)\oplus s_G(w)). \label{jg}
\end{align}
For decoding,
it first uses the decoder of the GS-BIS model to estimate the pair
($\widehat{w},\widehat{s_G(w)}$) and afterward the chosen secret key is retrieved by
\begin{align}
\widehat{s_C(w)} = s_C(\widehat{w})\oplus s_G(\widehat{w})\ominus \widehat{s_G(i)}, \label{sc+sc}
\end{align}
where $\oplus$ and $\ominus$ denote addition and subtraction modulo $M_S$.
This technique is also used in \cite{itw1}, \cite{itw3}, \cite{onur}, and so on.

\medskip
\noindent{\em Parameter Settings}:

First, we define $R_{J_G}$ and $R_{J_C}$ as the template rates in the GS-BIS and
the CS-BIS models encoders, respectively. Let $\delta$ be a small enough positive
and fix a block length $n$. We choose test channels $P_{U|Y}$ and $P_{V|U}$. Next,
%We select any pair of $(R_I,R_S)$ satisfying $0 \leq R_I + R_S < I(Z;U)$,
%$R_I < [I(Y;V) - I(Y;U) + I(Z;U)]^{+}$, and $R_S < H(Y|V)$.
We set $R_I = I(Z;V)-\delta$, $R_S = I(Z;U) - I(Z;V) - \delta$,
$R_{J_C} = I(Y;U) + \delta$, and $R_L = I(X;U) - I(Z;U) + I(Z;V) + 2\delta$.
%\footnote{\scriptsize{{Setting $R_I$ under these conditions, $\alpha$ or $R_S$ is accordingly smaller
%than $H(Y|V)$.}}} 
We also set the number of individuals $M_I = 2^{nR_I}$, the number of secret key
$M_S = 2^{nR_S}$, and the number of templates $M_{J_C} = 2^{nR_{J_C}}$ for the CS-BIS encoder 
and $M_{J_G}=\frac{M_{J_C}}{M_S}=
2^{n(I(Y;U) - I(Z;U) + I(Z;V) + 2\delta)}$ for the GS-BIS encoder, respectively.
%\renewcommand\thefootnote{1}
%\footnotetext{This study was supported in part by JSPS and NUS under 
%the Japan-Singapore Research Cooperative Program.}
%\renewcommand\thefootnote{\arabic{footnote}}
%We generate
%templates $M_U$ from $P_U$ $2^{n(I(U;Y)+2\delta)}$ pieces.
%It is obvious that $M_U$ is
%the product of $M_J$ and $M_S$.
%We also set $M_I = 2^{nR_I}$, $M_S = 2^{nR_S}$, and $M_J = 2^{nR_J}$.

\medskip
\noindent{\em Random Code Generation}:

Sequences $v^n_{m}$ are generated i.i.d.\ from $P_V$ for {$m~{\in}~[1,N_V]$,
where $N_V = 2^{n\left(I(Y;V) + \delta\right)}$.
%Function $\tilde{j}(m)$ is assigned to $v_m^n$ randomly and uniformly from the set $[1,M_J]$.
For each $m$, sequences $u^n_{k|m}$ are generated from {the memoryless channel}
$P_{U^n|V^n=v_{m}^n}$ for $k~{\in}~[1,N_U]$, where $N_U = 2^{n\left(I(Y;U|V) + \delta\right)}$.
{The indexes of these} $N_U$ codewords are
{permuted by {an uniformly distributed} permutation $\pi_m$ on $[1, N_U]$} and divided
equally into $N_B = 2^{n\left(I(Y;U|V) - I(Z;U|V) + 2\delta\right)}$ bins.
That is, the first bin contains $\{\tilde{u}^n_{1|m},\cdots,\tilde{u}^n_{M_S|m}\}$, the second bin
contains $\{\tilde{u}^n_{M_S+1|m},\cdots,\tilde{u}^n_{2M_S|m}\}$, and so on,
where ${\tilde{u}_{k|m}^n = u_{\pi_m^{-1}(k)|m}}$ denotes the $k$th codeword after the permutation.
Consequently, each bin contains $M_S$ codewords {in a} random order.
Bins are indexed by $b \in [1,N_B]$ and codewords inside a certain bin
are indexed by {$s \in \mathcal{S}$}.
Without loss of generality, there exists a one-to-one mapping between $k$ and
the pair $(b,s)$.

\medskip
\noindent{{\em Encoding (Enrollment)}}:

{When {the GS-BIS encoder}, used as a component inside the CS-BIS encoder,
observes the bio-data sequence $y^n_i \in \mathcal{Y}^n$,
{the} component {looks for $(m, k)$ such that
$(y^n_i,v^n_{m},u^n_{k|m}) \in A_{\epsilon}^{(n)}(YVU)$}.
In case there are more than one such pairs, the component picks one of them uniformly at random.
Assume that the component found a corresponding pair
$(m,k)$, denoted as $(m(i),k(i)) = (m(i), b(i), s(i))$,
satisfying the jointly typical condition above.
Then, the component sets $j_G(i) = (m(i), b(i))$ and
$s_G(i) = s(i)$ and shares them to the CS-BIS encoder.
After that, the CS-BIS encoder uses {$s_G(i)$} to mask the chosen secret
$s_C(i)$ by $s_C(i)\oplus s_G(i)$. This masked information is combined with
$j_G(i)$ to form the template $j_C(i)$ as
$j_C(i) = \left(j_G(i),s_C(i)\oplus s_G(i)\right) = \left(m(i), b(i),s_C(i)\oplus s_G(i)\right)$.
The template is stored at position $i$ in the database.}
If there do not exist such $m$ and $k$, the component shares $j_G(i) = (1,1)$ and $s_G(i) = 1$
to the CS-BIS encoder. In this case, the CS-BIS encoder declares error.
%and
%$s(i)$ is handed back to individual $i$.
%In case there do
%not exist such $m$ and $k$ {or there are multiple pairs of $m$ and $k$}
%, we set $j(i) = \tilde{j}(1)$ {and} $s(i) = \tilde{s}(1,1)$.
%We set $(j(i),s(i)) = f(y^n_i)$.
%Moreover, the encoder checks if there is another
%$\tilde{y}^n \in A_{\epsilon}^{(n)}(Y)$ such that $l_s(\tilde{y}^n) = s(i)$
%and $l_h(\tilde{y}^n) = h(i)$. If not, the template $h(i)$ are stored
%at location $i$ and the secret $s(i)$ is handed over to the individual.
%If such an $\tilde{y}^n$ exists, however, the location stays
%empty and no secret is offered to the individual.

\medskip
\noindent{{\em Decoding (Identification)}}:

The GS-BIS {decoder}, embedded as a component inside the CS-BIS decoder,
has access to {all records} in the database $\{\left(m(1), b(1),s_C(1)\oplus s_G(1)\right),
\cdots, (m(M_I), b(M_I), s_C(M_I)\oplus s_G(M_I))\}$
(the CS-BIS decoder also can).
When the component receives $z^n$
(the noisy version of identified individual sequence $x^n_w$),
it checks if the codeword pair
$(v^n_{m(i)},u^n_{b(i),s|m(i)})$ is jointly typical with $z^n$ for all
$i \in \mathcal{I}$ {with} some
$s \in \mathcal{S}$, i.e.
$(z^n,v^n_{m(i)},u^n_{b(i),s|m(i)}) \in A_{\epsilon}^{(n)}(ZVU)$}.
{If there exists a unique pair ${(i,s)}$ for which this condition holds,
then the component sets $(\widehat{w},\widehat{s_G(w)})=(i,s)$ and
forwards the pair $(\widehat{w},\widehat{s_G(w)})$
to the CS-BIS decoder. After getting it,
the CS-BIS decoder outputs $\widehat{w} = i$ and $\widehat{s_C(w)}$ as
the result of $s_C(\widehat{w})\oplus s_G(\widehat{w})\ominus \widehat{s_G(w)}$.
%$(\widehat{w},\widehat{s(w)}) = {(i,s)}$
%as the index of template where} $j(\widehat{w}) = m$ in the database
%as the estimated index and secret key, respectively.
Otherwise, the component shares {the index of the template $(1,1)$
and $\widehat{s_G(w)}=1$} to the CS-BIS decoder.
Upon detecting these information, the CS-BIS decoder declares error.
%There are three possible error events happen at the decoder
%{(i)} there does not exist such a pair $(i,s)$,
%{(ii)} {such a pair $(i,s)$ exists} but there are some
%$s' \neq {s}$ ($s' \in \mathcal{S}$) such that}
%$(z^n,v^n_{{m(i)}},
%u^n_{{b(i)},s'|{m(i)}}) \in A_{\epsilon}^{(n)}(ZVU)$ satisfies, or
%{(iii)} {such a pair $(i,s)$ exists but} there are some $i' \neq {i}$ such that the pair
%$(v^n_{m(i')},u^n_{b(i'),{s'}|m(i')})$
%is jointly typical with $z^n$ for some {$s' \in \mathcal{S}$}.
%subtraction modulo $M_S$between the masked information and the estimated $\widehat{s_G(w)}$.

Next we check that the conditions of (\ref{a})--(\ref{shiki4})
in Definition \ref{def1} averaged over randomly chosen codebook $\mathcal{C}_n$,
which is defined as the set $\{V_{m}^n, U_{k|m}^n, {\Pi_m}:
m \in [1,N_V], k \in [1,N_U]\}$, {where $\Pi_m$ denotes the RV corresponding to
{the permutation of the indexes of the sequences} $\{U_{1|m}^n,\cdots, U_{N_U|m}^n\}$ for given $m$}.

\medskip
\noindent{{\em Analysis of Error Probability}}:

For individual $W=w$, the operation at the decoder (\ref{sc+sc}) means that
$\widehat{S_C(w)} = S_C(w)$ only if $\widehat{S_G(w)} = S_G(w)$.
It is shown that the error probability of individual $w$ for the GS-BIS model can
be made that
$
\Pr\{(\widehat{W},\widehat{S_G(W)}) \neq (W,S_G(W))|W = w\} \leq 4\delta.
$
The {detailed proof} is provided in \cite[Proof of Theorem 1]{vy3}.

Therefore, it follows that the error probability of individual $w$ for
the CS-BIS model can {also} be bounded by
\begin{align}
\Pr\{(\widehat{W},\widehat{S_C(W)}) \neq (W,S_C(W))|W = w\} \leq 4\delta \label{error}
\end{align}
for large enough $n$.

\medskip
\noindent{{\em Analyses of {Identification, Secrecy, and Template Rates}}}:

It is easy to confirm that (\ref{b}), (\ref{d}), and (\ref{c}) hold from
the parameter settings.
%{Similarly,} it is obvious that (\ref{c}) holds.

\medskip
\noindent{{\em Analysis of Privacy-Leakage {Rate}}}:

{It is shown in \cite[Appendix B-A]{itw3} that}
%In \cite{itw3}, Ignatenko and Willems also provided an observation on privacy-leakage of the two models that
\begin{align}
{I}(X^n_i;J_C(i)|\mathcal{C}_n)&= I(X^n_i;J_G(i),S_C(i)\oplus S_G(i)|\mathcal{C}_n) \nonumber \\
&\leq {I}(X^n_i;J_G(i)|\mathcal{C}_n). \label{xjc}
\end{align}
By using a result shown in \cite{vy3}, the privacy-leakage of the GS-BIS model can be bounded by
$
\textstyle \frac{1}{n}{I}(X^n_i;J_G(i)|\mathcal{C}_n) \leq {I(X;U) - I(Z;U) + I(Z;V) + 3\delta}
$
for large enough $n$. The detail proof is provided in \cite[Appendix B-C]{vy3}.
Then, the privacy-leakage of the CS-BIS model can also be made that
\begin{align}
\frac{1}{n}{I}(X^n_i;J_C(i)|\mathcal{C}_n) &\leq I(X;U) - I(Z;U) + I(Z;V) + 3\delta \nonumber \\
&= R_L + \delta \label{1_1}
\end{align}
for large enough $n$.

\medskip
\noindent{{\em Analysis of Secrecy-Leakage}}:

{We invoke the following} relation on secrecy-leakage between the CS-BIS and the
GS-BIS models \cite[Appendix B-A]{itw3}:
\begin{align}
\frac{1}{n}&I(J_C(i);S_C(i)|\mathcal{C}_n) \nonumber \\
&= \frac{1}{n}I(J_G(i),S_C(i)\oplus S_G(i);S_C(i)|\mathcal{C}_n) \nonumber \\
&\leq \frac{1}{n}\log M_S - \frac{1}{n}H(S_G(i)) + \frac{1}{n}I(J_G(i);S_G(i)|\mathcal{C}_n). \label{jcsc111}
\end{align}
In \cite[Appendix B-B]{vy3} and \cite[Appendix B-C]{vy3}, it is shown that
\begin{align}
\frac{1}{n}H(S_G(i)) &\ge \log M_S - 2\delta, \label{sc} \\
\frac{1}{n}I(J_G(i);S_G(i)|\mathcal{C}_n) &\leq 2\delta \label{jcsc}
\end{align}
for large enough $n$.

Substituting (\ref{sc}) and (\ref{jcsc}) into (\ref{jcsc111}), the secrecy-leakage
of the CS-BIS model is bounded by
\begin{align}
\frac{1}{n}I(J_C(i);S_C(i)|\mathcal{C}_n)\leq 4\delta \label{IJSc}
\end{align}
for large enough $n$.

Finally, by applying the selection lemma \cite[Lemma 2.2]{BB} to above results,
there exists at least
a good codebook satisfying all conditions in Definition \ref{def1} for large enough $n$.
\qed

\subsection{Converse Part}

We consider a more relaxed case where identified individual index $W$
is {\em uniformly} distributed over $\mathcal{I}$ and (\ref{a}) is replaced with
the average {error criterion}
%\vspace{-1mm}
\begin{align}
\Pr\{({\widehat{W}},\widehat{S(W)})\neq (W,S(W))\} \leq \delta. \label{average}
\end{align}
%\vspace{-5mm}
{We} shall show that the capacity region, which is not smaller than the original one {$\mathcal{R}$},
is contained in the right-hand side of
(\ref{theorem2}).

We assume that a rate tuple $(R_I,R_S,R_J,R_L)$ is achievable so that
there exists a pair of encoder and decoder $(f,g)$ such that all conditions in
Definition \ref{def1} with replacing (\ref{a}) by (\ref{average}) are satisfied for any
$\delta > 0$ and large enough $n$. 

Here, we provide {other key} lemmas used in this part.
For $t \in [1,n]$, we define auxiliary RVs $U_t$ and $V_t$ as
$
U_t = (Z^{t-1},J(W),S(W))
$
and
$
V_t = (Z^{t-1},J(W))
$, respectively. We denote a sequence of RVs
$Y^n_W  = (Y_1(W) ,\cdots,Y_n(W))$.

\begin{lemma} \label{markov122}

{The following Markov chain holds}
\begin{align}
Z^{t-1}-(Y^{t-1}(W),J(W),S(W))-Y_t(W). \label{zyyt}
%Z^{t-1}-(X^{t-1}(W),J(W),S(W))-X_t(W). \label{zxxt}
\end{align}
\end{lemma}
\noindent{{(Proof)}}~~~~{The proof is} given in \cite[Appendix C-A]{vy3}.
\qed

\begin{lemma} \label{leema}
There {exist} some RVs $U$ and $V$ which satisfy $Z-X-Y-U-V$ and
\begin{align}
%\sum_{\substack{ t=1 }}^nI(Z_{t};V_t) &= nI(Z;V), \label{ztvt} \\
%\sum_{\substack{ t=1 }}^nI(Z_{t};U_t) &= nI(Z;U), \label{ztut} \\
{\sum_{\substack{ t=1 }}^nI(Y_{t}(W);U_t)} &= nI(Y;U), \label{ytut} \\
\sum_{\substack{ t=1 }}^nI(Y_{t}(W);V_t) &= nI(Y;V). \label{ytvt}
%\sum_{\substack{ t=1 }}^nI(X_{t}(W);U_t) &= nI(X;U). \label{xtut}
%\sum_{\substack{ t=1 }}^nI(X_{t}(W);V_t) &= nI(X;V). \label{xtvt} 
\end{align}
\end{lemma}

\noindent{(Proof)}~~~~{The proofs are} provided in \cite[Appendix C-B]{vy3}.
\qed

In the following arguments, we fix auxiliary RVs $U$ and $V$ specified in Lemma \ref{leema}.

\medskip
\noindent{{\em Analysis of Identification and Secrecy Rates}}:

It can be shown that
\begin{align}
R_I &\le I(Z;V) + \delta + \delta_n, \label{rilast}\\
R_S &\le I(Z;U)-I(Z;V) + 2\delta + {\delta_n}, \label{rslast}
\end{align}
where {$\delta_n = \frac{1}{n}(1+\delta \log M_IM_S)$} and {$\delta_n \downarrow 0$} as $n \rightarrow \infty$.

\noindent{T}he proofs can be done by similar arguments of the analysis of identification and secrecy rates
in the converse part of \cite[Proof of Theorem 1]{vy3}.

\medskip
\noindent{{\em Analysis of Template Rate}}:

From (\ref{d}), it holds that
\begin{align}
n(R_J + \delta) &\geq \log M_{J} \geq H(J(W)) \nonumber \\
%&= I(J(W);S(W),Y^n_W)\nonumber \\
%&= I(J(W);S(W)) + I(J(W);Y^n_W|S(W))\nonumber \\
&\geq I(J(W);Y^n_W|S(W)) \nonumber \\
&=H(Y^n_W|S(W)) - H(Y^n_W|J(W),S(W)) \nonumber \\
&\overset{\mathrm{(a)}}= \sum_{\substack{ t=1 }}^n \Big\{ H(Y_t(W)) \nonumber \\
&~~~- H(Y_t(W)|J(W),S(W),Y^{t-1}(W))\Big\} \nonumber \\
&\overset{\mathrm{(b)}}= \sum_{\substack{ t=1 }}^n \Big\{ H(Y_t(W)) \nonumber \\
&~~~- H(Y_t(W)|J(W),S(W),Y^{t-1}(W),Z^{t-1})\Big\} \nonumber \\
%&\overset{\mathrm{(b)}}\geq \sum_{\substack{ t=1 }}^n I(Y_t(W);J(W),S(W),Y^{t-1}(W))\nonumber \\
&\overset{\mathrm{(c)}}\geq \sum_{\substack{ t=1 }}^n I(Y_t(W);Z^{t-1},J(W),S(W))\nonumber \\
&= \sum_{\substack{ t=1 }}^n I(Y_t(W);U_t)) \overset{\mathrm{(d)}}= nI(Y;U),
\end{align}
where
\begin{enumerate}[label=(\alph*)]
	\setcounter{enumi}{0}
        \item holds because $S(W)$ is independent of $Y^n_W$ and each symbol of $Y^n_W$ is i.i.d.,
        \item is due to (\ref{zyyt}) in Lemma \ref{markov122},
        \item follows because conditioning reduces entropy,
        \item holds due to (\ref{ytut}) in Lemma \ref{leema}.
\end{enumerate}
Thus, we obtain
\begin{align}
R_J \ge I(Y;U) - \delta. \label{rjjj}
\end{align}

\medskip
\noindent{{\em Analysis of Privacy-Leakage {Rate}}}:

It can be proved that
\begin{align}
R_L + \delta \geq I(X;U) - I(Z;U) + I(Z;V) - {\delta_n}. \label{rl}
\end{align}
\noindent{F}or detailed proof, the readers should refer to the analysis of
privacy-leakage rate in the converse part of \cite[Proof of Theorem 1]{vy3}
since similar approach is taken.

\medskip

By letting $n \rightarrow \infty$ and $\delta \downarrow 0$,
we obtain that the capacity region is contained in the right-hand side of (\ref{theorem1})
from (\ref{rilast}), \eqref{rslast}, \eqref{rjjj}, and (\ref{rl}).

To derive the bound on the cardinality of alphabet $\mathcal{U}$ in the region $\mathcal{A}_1$
{(cf.\ (\ref{theorem1}))},
we use the support lemma in \cite[Appendix C]{GK} to show that RV $U$ should have $|\mathcal{Y}|-1$
elements to preserve $P_{Y}$ and add three more elements to preserve $H(Z|U)$, $H(Y|U)$, and $H(X|U)$.
This implies that it suffices to take $|\mathcal{U}| \le |\mathcal{Y}|+2$
for preserving $\mathcal{A}_1$.
Similarly, for bounding the cardinalities of alphabets $\mathcal{U}$ and $\mathcal{V}$
in the region $\mathcal{A}_2$ {(cf.\ (\ref{theorem2}))}, we also utilize the same lemma to show that
$|\mathcal{V}| \le |\mathcal{Y}|+3$ and $|\mathcal{U}| \le (|\mathcal{Y}| + 2)(|\mathcal{Y}| + 3)$
suffice to preserve the regions $\mathcal{A}_1$ and $\mathcal{A}_2$.
\qed

\section{Conclusion and Future Works} \label{sec5}
In this study, we characterized the capacity region of identification, secrecy, template, and
privacy-leakage rates in the CS-BIS model. Compared to the model proposed in \cite{itw3}, we imposed
a noisy channel in the enrollment phase as seen in \cite{willems}, \cite{tuncel}, \cite{onur}, \cite{vy3}
and assumed that the prior distribution of the identified individual is unknown.
As special cases, the characterization
reduces to the result given by Ignatenko and Willems \cite[Theorem 2]{itw3} when the enrollment channel is noiseless
and there is no constraint on the template rate,
and also {matches with} the one given by G\"unl\"u and Kramer \cite[Theorem 2]{onur} where
there is only one individual.
For future work, we plan to analyze the capacity regions of the GS-BIS and CS-BIS
models under strong secrecy criterion in terms of secrecy-leakage.
%from verify if Theorem \ref{th1} can
%be re-characterized by only one auxiliary RV as is shown in \cite{onur}, \cite{itw1}, and \cite{lhp}
%in a slightly different scenario.
%\vspace{-1mm}

\end{document}